\RequirePackage{arydshln}
\documentclass[floatfix,aps,twocolumn,nofootinbib,superscriptaddress,preprintnumbers,pra,10pt]{revtex4-1}

\usepackage[utf8]{inputenc}
\usepackage{float}
\usepackage{colortbl}
\usepackage{amsmath,amssymb,amsfonts,bm,bbm,slashed}
\usepackage{dsfont} 
\usepackage{hyperref}
\usepackage{graphicx}
\usepackage{enumitem}
\usepackage{arydshln}
\usepackage{mathtools}
\usepackage{bbold}
\usepackage{braket}
\usepackage{multirow}
\usepackage{soul}
\usepackage{xcolor}
\usepackage{soul}

\makeatletter
\g@addto@macro\bfseries{\boldmath}
\makeatother

\newcommand{\gsim}{\lower.7ex\hbox{$\;\stackrel{\textstyle>}{\sim}\;$}}
\newcommand{\lsim}{\lower.7ex\hbox{$\;\stackrel{\textstyle<}{\sim}\;$}}
\newcommand{\mui}{ m^{\scriptscriptstyle (i)}_u}

\newcommand{\mRi}{ m^{\scriptscriptstyle (i)}_R}
\newcommand{\mRt}{ m^{\scriptscriptstyle (3)}_R}
\newcommand{\mDi}{ m^{\scriptscriptstyle (i)}_D}
\newcommand{\mDt}{ m^{\scriptscriptstyle (3)}_D}
\newcommand{\mnui}{ m^{\scriptscriptstyle (i)}_\nu}
\newcommand{\mdi}{ m^{\scriptscriptstyle (i)}_d}
\newcommand{\mei}{ m^{\scriptscriptstyle (i)}_e}
\newcommand{\cO}{\mathcal{O}}
\newcommand{\cG}{\mathcal{G}}

\allowdisplaybreaks

\begin{document}

\preprint{MITP/20-083, ZU-TH-56/20}

\title{Flavor Non-universal Pati-Salam Unification and Neutrino Masses}
 
\author{Javier Fuentes-Mart\'{\i}n}
\email{jfuentes@uni-mainz.de}
\affiliation{PRISMA+ Cluster of Excellence \& Mainz Institute for Theoretical Physics, Johannes Gutenberg University, 55099 Mainz, Germany}
\author{Gino Isidori}
\email{isidori@physik.uzh.ch}
\affiliation{Physik-Institut, Universit\"at Z\"urich, CH-8057 Z\"urich, Switzerland}
\author{Julie Pag\`es}
\email{julie.pages@physik.uzh.ch}
\affiliation{Physik-Institut, Universit\"at Z\"urich, CH-8057 Z\"urich, Switzerland}
\author{Ben A. Stefanek}
\email{bestef@physik.uzh.ch}
\affiliation{Physik-Institut, Universit\"at Z\"urich, CH-8057 Z\"urich, Switzerland}

\begin{abstract}
\vspace{5mm}
We analyze the neutrino mass spectrum and discuss the extra-dimensional interpretation of a three-site Pati-Salam model which~i) unifies all families of quark and leptons,~ii)~provides a natural description of the Standard Model Yukawa couplings, iii)~could account for the recent $B$-physics anomalies.  The key feature of the model is a breaking of the Pati-Salam and electroweak gauge symmetries localized on opposite sites, communicated to the other sites in an attenuated manner via nearest-neighbor interactions. We show that in this context gauge-singlet fermions localized on each site, receiving hierarchical Majorana masses, can allow the implementation of an inverse seesaw mechanism leading to light anarchic neutrino masses consistent with data. The continuum limit of this three-site setup has a natural interpretation in terms of a warped extra dimension with three defects, where the required exponential hierarchies can be achieved from $\mathcal{O}(1)$ differences in the bulk field masses.
\vspace{3mm}
\end{abstract}

\maketitle

\section{Introduction}\label{sec:intro}
Quark-lepton unification in the style of Pati and Salam~\cite{Pati:1974yy} is a compelling ultraviolet extension of the Standard Model (SM). In addition to a simplification of the SM matter content, as quarks and leptons are unified into the fundamental representation of $SU(4)$, the non-Abelian nature of the Pati-Salam (PS) gauge group predicts the quantization of hypercharge. Furthermore, due to baryon number arising as an accidental global symmetry at the renormalizable level, proton stability is a prediction of PS-style unification models. These features on their own are enough to beg the question of whether the quark-lepton unification scale can be as low as TeV, which is the scale currently being probed by the Large Hadron Collider (LHC). Making this question even more timely is that PS-style quark-lepton unification predicts the existence of a gauge vector leptoquark (LQ) transforming as  $U_{1}^{\mu} = ({\bf 3},{\bf 1},2/3)$ under the SM gauge group, which has been shown to be the ideal single mediator explanation to address the recent $B$-anomalies if its mass is no more than a few TeV~\cite{Alonso:2015sja,Calibbi:2015kma,Barbieri:2015yvd,Bhattacharya:2016mcc,Buttazzo:2017ixm,DiLuzio:2017vat}.

There are two major issues with flavor-universal PS unification at the TeV scale. The first is that in order to satisfy stringent bounds on flavor changing neutral current (FCNC) processes involving the light SM fermions, the mass of a flavor-blind PS leptoquark must be pushed above $10^{3}$ TeV~\cite{Valencia:1994cj,Smirnov:2007hv,Kuznetsov:2012ai,Giudice:2014tma,Smirnov:2018ske}. The second is that up-type quarks and neutrinos are unified within the same ${\bf 4}$ of $SU(4)$, leading to the phenomenologically disastrous prediction $\mui = \mnui$. In models where the PS gauge group is broken at the grand-unification scale, this problem can be naturally solved in the context of a Type-1 seesaw mechanism~\cite{Babu:1992ia,Minkowski:1977sc,Mohapatra:1979ia,Yanagida:1979as,GellMann:1980vs}, but this solution fails dramatically for low-scale PS unification.

The simplest way of addressing the first issue is to implement PS unification in a flavor non-universal 
manner~\cite{Bordone:2017bld,Greljo:2018tuh,Fuentes-Martin:2020bnh}.  
Interestingly, third-family quark-lepton unification close to the electroweak (EW) scale is phenomenologically allowed.
Furthermore, this setup naturally accommodates an accidental $U(2)^5$ global flavor symmetry at the TeV-scale: a key ingredient required to evade the tight bounds on FCNCs while simultaneously addressing the $B$-anomalies~\cite{Barbieri:2015yvd,Buttazzo:2017ixm,Cornella:2019hct,Fuentes-Martin:2019mun}.

Most of the existing TeV-scale unification models only attempt third-family PS unification, keeping the light families SM-like~\cite{Greljo:2018tuh,Fuentes-Martin:2020bnh}. An exception to this is the ${\rm PS}^3$ model~\cite{Bordone:2017bld}, which postulates a copy of the PS gauge group for every generation, yielding family-by-family unification of quarks and leptons at different energy scales. In particular, the third family is unified at the TeV scale, while the light families are unified at significantly higher scales in order to evade the FCNC bounds.
A key feature of this construction is that, to first approximation, the Higgs field responsible for the electroweak symmetry breaking (EWSB) 
is charged only under the third family (or third site) PS gauge group. This implies that EWSB is communicated to the light families via  
the effective mixing of this field with heavier states, charged under the different gauge groups. This mixing 
is suppressed by the scale hierarchy (or the {\em separation} between sites), naturally leading to the 
hierarchical flavor structure of the SM. 
This construction finds a natural justification in 
the context of a higher-dimensional theory, where the generation index is in one-to-one relation to the
location of four-dimensional (4D) branes along a hypothetical compact fifth dimension. 

The  ${\rm PS}^{3}$ setup forces us to address three separate neutrino mass problems at the three different energy scales.
The neutrino mass problem in the context of a low-scale flavor-universal PS model was first solved in~\cite{Perez:2013osa} by a simple extension of the matter content: gauge-singlet fermions were introduced to implement the
so-called \emph{inverse seesaw} (ISS) mechanism~\cite{Mohapatra:1986aw,Mohapatra:1986bd,GonzalezGarcia:1988rw}. The resulting neutrino masses scale parametrically as $m_{\nu} \sim \mu\, (m_u/m_R)^{2} $, where
$m_R$ is of order the PS unification scale and $\mu$ is the Majorana mass of the new singlets.
The latter is the only fermion number violating parameter of the theory and can therefore be taken very small, providing a natural 
 justification for the smallness of neutrino masses.
 It was recently shown that the same mechanism can also be implemented to solve the neutrino mass problem of a 
 flavor non-universal PS model, with third-family unification at the TeV scale~\cite{Greljo:2018tuh}. 
 There, however,  the required  flavor structure to obtain anarchical neutrino masses and mixings is not explained but rather put by hand. It is therefore interesting to ask whether a solution of the neutrino mass problem in the three-site ${\rm PS}^{3}$ model might also result in an explanation of this structure, somehow analogously to what happens for the flavor structure of SM Yukawa couplings. 

At first glance, obtaining anarchical neutrino masses and mixings from a ${\rm PS}^{3}$-type construction seems difficult as the family-by-family ratios $(\mui/ \mRi)^{2}$ are extremely hierarchical. This is because the quark masses and 
PS-breaking $\nu_R$ masses are inversely hierarchical, with the highest $\mRi$ corresponding to the lowest $\mui$, 
and vice-versa. Indeed, if one implements the ISS mechanism by adding three gauge-singlet fermions, one obtains an extremely hierarchical active neutrino mass spectrum if the Majorana mass for all the singlets originates from the same scale ($\mu_{i} \sim \mu$). 

In this letter, we show that an anarchical neutrino flavor structure can be naturally realized in ${\rm PS}^{3}$ if the Majorana masses $\mu_i$ arise dynamically from the vacuum expectation value (VEV) of a SM-singlet scalar field. The breaking of the fermion number symmetry is then assumed to be localized in the first family (or first site) and, as in the case of EWSB, communicated to the other sites only via small nearest-neighbor interactions, yielding  $m_{\nu}^{\scriptscriptstyle (i)} \sim \mu_{i} (\mui / \mRi )^{2} \approx\mathrm{const}$. This amounts to a site-by-site solution to the neutrino mass problem where the hierarchy in $(\mui/ \mRi)^{2}$ is compensated by the corresponding inverse hierarchy in $\mu_{i}$.

Besides addressing the problem of neutrino masses, we further discuss how the discrete multi-scale structure of the ${\rm PS}^{3}$ framework could be justified within a continuous five-dimensional (5D) UV theory.\footnote{Alternative UV models based on variations of 
the original PS model, aimed at addressing the $B$-physics anomalies, have been proposed in~\cite{Blanke:2018sro,Fornal:2018dqn}.}
As we show, the flavor structure of  the model can naturally be understood as originating from scalar fields taking exponentially decaying VEVs in the bulk. In particular, the breaking of the fermion number and PS symmetries are assumed to originate at the same boundary of a compact extra dimension, while EWSB originates at the opposite boundary. This way, the scalar profiles evaluated on three sites (or branes) in the bulk generate the required hierarchies from $\cO(1)$ differences in the scalar bulk masses. Moreover, as we discuss in detail, the requirement of addressing the $B$ anomalies, points toward a warped geometry for the extra dimension.

The structure of the paper is the following: in Section~\ref{sec:PS3model}, we briefly recall the PS$^3$ model, introducing 
a few modifications with respect to the original formulation~\cite{Bordone:2017bld} in order to facilitate its interpretation 
within the context of a higher-dimensional theory. In Section~\ref{sec:nu}, we show how to extend the 
model to implement the ISS mechanism and address the phenomenology of the neutrino sector.
In Section~\ref{sec:5D}, we discuss the possible embedding of the model into a continuous 5D description. 
The results are summarized in Section~\ref{sec:conclusions}. 

\section{The PS$^3$ model}\label{sec:PS3model}
\label{sec:PS3rev}

The ${\rm PS}^3$ model introduced in~\cite{Bordone:2017bld} consists 
of a 4D (de)construction~\cite{ArkaniHamed:2001ca,Hill:2000mu,Cheng:2001vd} of the original 
PS model~\cite{Pati:1974yy}  into three sites. The 4D gauge sector is\footnote{We assume that the ${\rm PS}_1$ factor is reduced to ${\rm PS}_1^\prime\equiv [SU(4)\times SU(2)_L\times U(1)_R]_1$ before inflation, avoiding the monopole problem of low-scale ${\rm PS}$ models~\cite{Jeannerot:2000sv,Greljo:2019xan}. This can be easily achieved by an appropriate $SU(2)_R$ breaking source, without affecting our discussion.}
\begin{align}
{\rm PS}^3\equiv{\rm PS}_1\times{\rm PS}_2\times{\rm PS}_3\,,
\end{align}
where ${\rm PS}\equiv SU(4)\times SU(2)_L\times SU(2)_R$.
Each  ${\rm PS}_i$ group acts, separately, on each of the SM fermion families, described by the fields
\begin{align}
\Psi_L^{\scriptscriptstyle (i)}\sim (\mathbf{4},\mathbf{2},\mathbf{1})_i\,,\qquad\Psi_R^{\scriptscriptstyle (i)}\sim (\mathbf{4},\mathbf{1},\mathbf{2})_i\,,
\end{align}
with the inclusion of three right-handed neutrinos.

\begin{figure}[t]
\centering
\includegraphics[scale=0.32]{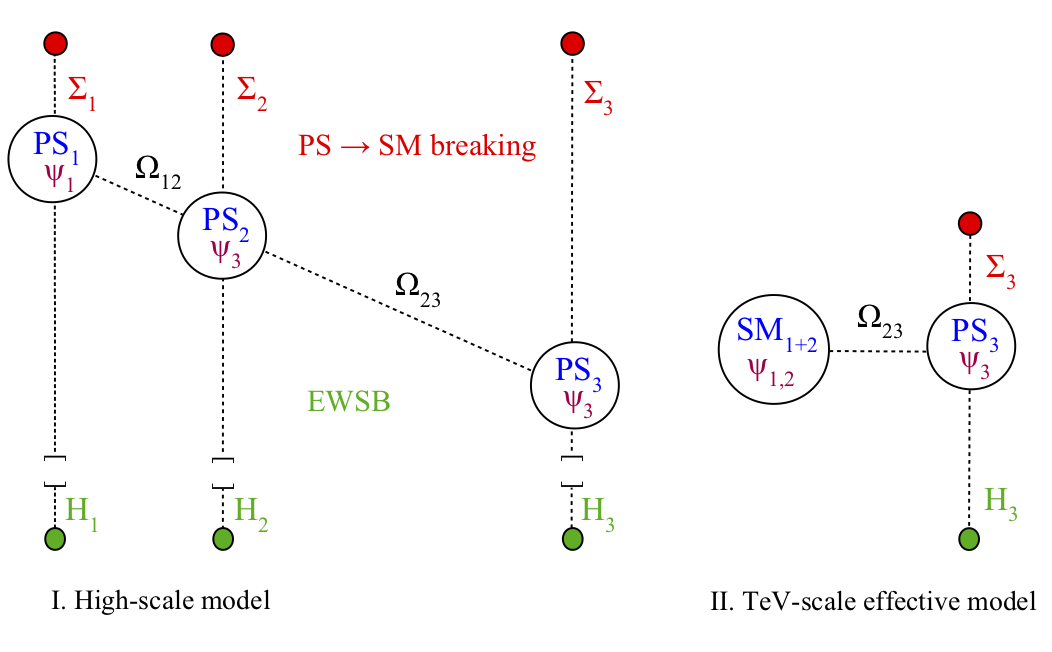}
\caption{Schematic representation of the SSB structure in ${\rm PS}^3$: each dotted line denotes a field (or set of fields) with non-vanishing VEV. Long/short lines qualitatively indicate small/large VEVs.}
\label{fig:PS3breaking}
\end{figure}

The breaking of the ${\rm PS}^3$ symmetry down to the (flavor-universal) SM group is realized via a series of spontaneous 
symmetry breaking (SSB) steps, summarized in Fig.~\ref{fig:PS3breaking}. Each of these breakings belongs to one of the following categories: 
\begin{itemize}
\item[i)] \textit{Horizontal breaking}, where (part of) the symmetries of two adjacent sites are broken to their diagonal subgroup.  This is achieved by a set of non-linear scalars or \textit{link fields}, transforming in bi-fundamental representations of ${\rm PS}^3$,
\begin{align}
\begin{aligned}
\Omega_{ij}^4\sim  (\mathbf{4},\mathbf{1},\mathbf{1})_i\times  (\mathbf{\bar 4},\mathbf{1},\mathbf{1})_j\,,\\
\Omega_{ij}^L\sim  (\mathbf{1},\mathbf{2},\mathbf{1})_i\times  (\mathbf{1},\mathbf{\bar 2},\mathbf{1})_j\,,\\
\Omega_{ij}^R\sim  (\mathbf{1},\mathbf{1},\mathbf{2})_i\times  (\mathbf{1},\mathbf{1},\mathbf{\bar 2})_j\,,
\label{eq:linkfields}
\end{aligned}
\end{align}
that acquire a VEV at the scale $f_{ij}^{4,L,R}$. Note that this is different from the original implementation in~\cite{Bordone:2017bld}, where the links fields were introduced as linear fields. Having non-linear fields implies that the model has a UV cut-off at $\Lambda\sim 4\pi f_{ij}^{4,L,R}$ at which unitarity is lost. Unitarity could be restored by replacing the links by linear fields, as in~\cite{Bordone:2017bld}, or by an infinite tower of heavy resonances, as in an extra-dimensional (or composite) description.

\item[ii)] \textit{Vertical breaking}, where the symmetry of a given site is broken into a smaller subgroup. We consider two types of vertical breakings:
\begin{itemize}
\item[ii.a)] ${\rm PS}_i\to {\rm SM}_i$ breaking: the PS symmetry of a given site is broken to its SM subgroup 
by the VEV of the scalar fields $\Sigma_i\sim (\mathbf{4},\mathbf{1},\mathbf{2})_i$.\footnote{Another relevant scalar field, whose VEV leaves the SM subgroup invariant, is $\Sigma_{15}\sim(\mathbf{15},\mathbf{1},\mathbf{1})$. This field helps splitting quark and lepton masses and introduces new sources of flavor violation in the leptoquark interactions~\cite{DiLuzio:2018zxy,Cornella:2019hct}. For simplicity, we will not discuss this field here.} To comply with the stringent flavor constraints on the PS leptoquark, this breaking is assumed to occur mainly in the first site. More precisely, we assume 
$\langle \Sigma_3\rangle=\epsilon^\Sigma_{23}\,\langle \Sigma_2\rangle=\epsilon^\Sigma_{23}\,\epsilon^\Sigma_{12}\,\langle \Sigma_1\rangle$, with $\epsilon^\Sigma_{ij}$ being small parameters that determine the site-by-site attenuation of the $\Sigma_1$ VEV.
\item[ii.b)] EWSB: The EW subgroup [$SU(2)_L\times U(1)_Y]_{i}$ of a given site is broken to the corresponding QED-like subgroup 
by the VEV of a Higgs-like field $H_i\sim(\mathbf{1},\mathbf{2},\mathbf{\bar 2})_i$. Contrary to the previous case, in order to explain the 
hierarchical structure of the SM Yukawa couplings, the Higgs VEV is assumed  
to be dominantly localized on the third site. More precisely, we assume $\langle H_1\rangle=\epsilon^H_{12}\,\langle H_2\rangle=\epsilon^H_{12}\,\epsilon^H_{23}\,\langle H_3\rangle$, with the attenuation parameters $\epsilon^H_{ij}$ taken to be small. 
\end{itemize}
Within a renormalizable 4D description, the quasi-localized $\Sigma_i$ and $H_i$ VEV structure can easily be obtained from a multi-site scalar potential with nearest-neighbor interactions~\cite{Cheng:2001vd,Allwicher:2020esa}.  In this construction, the $\epsilon^{H,\Sigma}_{ij}$ parameters are related to ratios of masses of the different scalar fields. As we discuss in Section~\ref{sec:5D}, this structure could originate from 5D scalar fields localized on opposite sides of a compact extra dimension with exponentially decaying VEV profiles.

This construction is different but in close analogy to the original ${\rm PS}^3$ 
proposal~\cite{Bordone:2017bld}, where the vertical breaking was assumed to be fully localized on the third (first)
site for $H~(\Sigma)$, and an effective small delocalization was achieved via higher-dimensional operators written in terms 
of the (linear) link fields. 
\end{itemize}

The different symmetry breaking scales are constrained by flavor-violating processes and by the observed Yukawa patterns. More precisely, the tight constraints from FCNCs fix the scale of ${\rm PS}_{1,2}\to {\rm SM}_{1,2}$ vertical breakings as well as 1-2 horizontal breakings to be above $10^3$~TeV. Below $\Lambda_{12}\approx 10^3$~TeV, the gauge symmetry is thus reduced to ${\rm SM}_{1+2}\times {\rm PS}_3$, and the model presents a global accidental $U(2)^5$ flavor symmetry acting on the first- and second-family fermions. The $U(2)^5$ symmetry ensures enough protection against flavor constraints, so that the VEV of $\Sigma_3$ and the 2-3 link fields (and hence the breaking of ${\rm PS}_3$) can be as low as few TeV~\cite{Cornella:2019hct,Fuentes-Martin:2019mun}.
On the other hand, as previously mentioned, the observed Yukawa patterns suggest that the dominant source of EWSB should be located on the third-family site. This way, the Yukawa interactions
\begin{align}\label{eq:YukDiag}
-\mathcal{L}_Y\supset y_i\,\bar \Psi_L^{\scriptscriptstyle (i)}H_i\Psi_R^{\scriptscriptstyle (i)}+\tilde y_i\,\bar \Psi_L^{\scriptscriptstyle (i)}\tilde H_i\Psi_R^{\scriptscriptstyle (i)}+{\rm h.c.}\,,
\end{align}
with $\tilde H_i=i\sigma_2\, H_i^*\, i\sigma_2$ denoting the conjugate Higgs field, are sufficient to describe the fermion masses with $\cO(1)$ Yukawa couplings.\footnote{The smallness of tau and bottom Yukawa could be explained by extending the Higgs sector to a type-II two-Higgs-doublet model with $v_u\gg v_d$.} To reproduce the Cabibbo--Kobayashi-Maskawa (CKM) matrix, we introduce nearest-neighbor interactions between fermions
\begin{align}\label{eq:YukMix}
\begin{aligned}
-\mathcal{L}_Y&\supset y_{23}\,\epsilon_{23}^L\,\bar \Psi_L^{\scriptscriptstyle (2)}\,\frac{\Omega^4_{23}}{f^4_{23}}\,\frac{\Omega_{23}^L}{f^L_{23}}\,H_3\Psi_R^{\scriptscriptstyle (3)}\\
&\quad+ y_{12}\,\epsilon_{12}^L\,\bar \Psi_L^{\scriptscriptstyle (1)}\,\frac{\Omega^4_{12}}{f^4_{12}}\,\frac{\Omega_{12}^L}{f^L_{12}}\,H_2\Psi_R^{\scriptscriptstyle (2)}+ {\rm h.c.}\,,
\end{aligned}
\end{align}
and analogous terms with the conjugate Higgs field. Since these correspond to interactions between fermions of different sites, we include the small parameters $\epsilon_{ij}^L$ which, similarly to the case of the $\Sigma_i$ and Higgs VEVs, determine the localization of the left-handed fermions in their corresponding sites. We further assume that right-handed fermions are fully localized, i.e. $\epsilon^{R}_{ij} = 0$. 
A renormalizable 4D description where the $\epsilon_{ij}^L$ can be computed in terms of scale ratios (and additional interaction terms)
can be obtained by introducing a series of vector-like fermions, charged under the 
subgroups~\cite{Cheng:2001vd,Bordone:2017bld,Greljo:2018tuh,Allwicher:2020esa}. 

The Yukawa Lagrangian in~\eqref{eq:YukDiag} and~\eqref{eq:YukMix} preserves the accidental $U(2)^5$ flavor symmetry of the gauge sector in the limit where the Higgs VEV and the left-handed fermions are fully localized, i.e. $\epsilon^H_{ij}, \epsilon_{ij}^L\to0$. These parameters therefore act as spurions of the flavor symmetry. Their size can be deduced from the CKM mixing matrix, the fermion mass ratios and the fit to the anomalies in semileptonic $B$ decays~\cite{Fuentes-Martin:2019mun}, suggesting the approximate relations
\begin{align}\label{eq:epsLandH}
\epsilon^H_{ij}\sim(\epsilon^L_{ij})^2 \sim 10^{-2}\,,\qquad\epsilon_{12}^L\approx 2\,\epsilon_{23}^L\,.
\end{align}

An appealing feature of this multi-site construction is that the postulated mass hierarchies, 
which are deduced by the flavor hierarchies and the absence of large FCNCs, 
are protected against large radiative corrections by the nearest-neighbor structure of the interactions~\cite{Allwicher:2020esa}. 
In particular, the Higgs sector is radiatively stable:  the light Higgs field on the third site, responsible for EWSB, is 
coupled to the high-scale fields on the first site only at high orders in the loop expansion. As a  result, the model suffers 
of a moderate fine-tuning problem, not different than any other model with TeV-scale dynamics~\cite{Allwicher:2020esa}. 

\section{Extension of the neutrino sector}
\label{sec:nu}

In the original formulation of the ${\rm PS}^3$ model, EWSB provides the only source for neutrino masses. Since quarks and leptons are unified into $SU(4)_i$ multiplets, this implies the mass relations $\mei= \mdi$ and $\mnui = \mui$ for $SU(4)_i$-singlet Higgses. While the first mass relation is approximately realized and only needs to be slightly perturbed, the second mass relation is clearly inconsistent with observations. Adding Higgs fields transforming in the $\mathbf{15}$ representation of $SU(4)_i$, as done in~\cite{Bordone:2017bld}, would modify this unwelcome prediction. However, the required tuning between the $SU(4)$-preserving and the $SU(4)$-violating Higgs VEVs is at the $10^{-12}$ level, making this solution unsatisfactory. 

On the other hand, right-handed neutrino masses can only be as high as the corresponding ${\rm PS}_i$-breaking scales, making normal seesaw mechanisms, like type-I seesaw, rather unnatural for a TeV-scale ${\rm PS}_3$ breaking. As pointed out in~\cite{Greljo:2018tuh}, a more natural implementation is achieved via an inverse seesaw mechanism~\cite{Mohapatra:1986aw,Mohapatra:1986bd,GonzalezGarcia:1988rw}.
The simplest extension of the ${\rm PS}^3$   set-up  yielding such mechanism consists in adding  three chiral gauge-singlet fermions $S_L^{\scriptscriptstyle (i)}$, one for each family,\footnote{Given the presence of three right-handed neutrinos in the model, 
this is the minimal viable ISS implementation~\cite{Abada:2014vea}.} responsible for the breaking of fermion number(s). 
The latter breaking can be made directly proportional to the observed neutrino masses by introducing appropriate couplings 
between $S_L$ and $\Psi_R$ (see below). The (arbitrarily small) breaking of fermion number can be obtained  
adding Majorana mass terms for the $S_L$.  As anticipated, the strong hierarchy in the Dirac masses and ${\rm PS}_i$ breaking scales 
calls for highly hierarchical Majorana masses to reproduce an anarchical light neutrino mass matrix. We can achieve this goal assuming that these masses arise from the VEVs of new gauge-singlet fields $\Phi_i$, responsible of the spontaneous breaking of fermion number. This way, the required hierarchies can be understood in terms of VEV localizations, analogously to the $\Sigma$ and $H$ cases.  As for $\Sigma$, we assume that the VEV of $\Phi$ is dominantly localized in the first site. Namely, we take  $\langle \Phi_3\rangle=\epsilon^\Phi_{23}\,\langle \Phi_2\rangle=\epsilon^\Phi_{23}\,\epsilon^\Phi_{12}\,\langle \Phi_1\rangle$, with $\epsilon^\Phi_{ij}\ll1$ controlling the site-by-site VEV attenuation. 

The relevant Lagrangian for neutrino masses thus reads
\begin{equation}  \label{eq:NeutrinoDiag}
-\mathcal{L} _\nu \supset  y^\Sigma_i\, \bar  S_{L}^{\scriptscriptstyle (i)}\, \Sigma_i^\dag\, \Psi_{R}^{\scriptscriptstyle (i)}  +  \frac{y^\Phi_i}{2}\, \bar S_{L}^{\scriptscriptstyle (i)}\, \Phi_i \,{S_{L}^{\scriptscriptstyle(i)}}^c + {\rm h.c.}\,,
\end{equation} 
which, combined with the Dirac neutrino masses from~\eqref{eq:YukDiag}, yields the inverse seesaw mechanism. Analogously to the Higgs Yukawa sector, we introduce nearest-neighbor interactions accounting for interfamily mixing
\begin{align} \label{eq:NeutrinoMix}
\begin{aligned}
-\mathcal{L}_\nu &\supset   y_{21}^\Sigma\,\epsilon^S_{21}\, \overline S_{L}^{\scriptscriptstyle (2)}\, \Sigma_1^\dag\, \Psi_{R}^{\scriptscriptstyle (1)} +y_{32}^\Sigma\,\epsilon^S_{32}\, \overline S_{L}^{\scriptscriptstyle (3)}\, \Sigma_2^\dag\, \Psi_{R}^{\scriptscriptstyle (2)}\\
&\quad+y_{31}^\Sigma\,\epsilon^S_{32}\,\epsilon^S_{21}\, \overline S_{L}^{\scriptscriptstyle (3)}\, \Sigma_1^\dag\, \Psi_{R}^{\scriptscriptstyle (1)}    
+ {\rm h.c.} \,,
\end{aligned}
\end{align}
where we have used the freedom to rotate away off-diagonal terms in the $\Phi$ interactions by an appropriate $S_L$ redefinition. Once more, the $\epsilon^S_{ij}$ parameters control the site localization of the $S_L^{\scriptscriptstyle (i)}$ fields. Analogous to the nearest neighbor interactions of the Higgs Yukawa sector in~\eqref{eq:YukMix}, these parameters can be understood as arising from the mixing between $S_L^{\scriptscriptstyle (i)}$ and SM-singlet vector-like fermions. However, note that contrary to~\eqref{eq:YukMix}, no link fields are present since $U(1)_F$ is a global symmetry.

\subsection{Neutrino masses and mixings}

After spontaneous symmetry breaking, we obtain the following light neutrino mass matrix (see Appendix~\ref{app:NeutrinoDetails} for details)
\begin{align}
m_\nu\approx m_D\, m_R^{-1}\,\mu\;(m_D\,m_R^{-1})^\intercal\,,
\end{align}
where the different mass matrices are defined as
\begin{align}
m_D&=\langle H_3\rangle
\begin{pmatrix}
y_1^\nu\epsilon^H_{12}\,\epsilon^H_{23} & y_{12}^\nu\,\epsilon^L_{12}\,\epsilon^H_{23} & y_{13}^{\nu}\epsilon^L_{12}\epsilon^L_{23}\\[2pt]
0 & y_2^\nu\epsilon^H_{23} & y_{23}^\nu\,\epsilon^L_{23}\\[2pt]
0 & 0 & y_3^\nu
\end{pmatrix}
\,, \nonumber \\[4pt]
m_R&=\langle \Sigma_1\rangle
\begin{pmatrix}
y_1^\Sigma & 0 & 0\\[2pt]
y_{21}^\Sigma\,\epsilon^S_{21} & y_2^\Sigma\,\epsilon^\Sigma_{12} & 0\\[2pt]
y_{31}^\Sigma\,\epsilon^S_{32}\,\epsilon^S_{21} & y_{32}^\Sigma\,\epsilon^S_{32}\,\epsilon^\Sigma_{12} & y_3^\Sigma\,\epsilon^\Sigma_{12}\,\epsilon^\Sigma_{23} 
\end{pmatrix}
\,,\nonumber \\[4pt]
\mu&=\langle\Phi_1\rangle\,\mathrm{diag}(y_1^\Phi,y_2^\Phi\epsilon^\Phi_{12},y_3^\Phi\epsilon^\Phi_{12}\,\epsilon^\Phi_{23})\,,
\label{eq:massMats}
\end{align}
with $m_{D}$ written in the charged lepton mass eigenbasis. Here, $y^\nu$ is the neutrino Yukawa coupling, which we allow to
differ from the up-quark Yukawa, $y^u$, by $\cO(1)$ terms. 
As previously mentioned, this difference can be achieved via appropriate $SU(4)$-breaking sources. 

Our goal is to obtain a non-hierarchical structure for $m_\nu$, as indicated by neutrino oscillation data (i.e.~neutrino mass differences and mixing angles) without tuning the $y_{ij}^\nu$ and $y_{ij}^\Sigma$ coefficients, which we assume to be of $\cO(1)$.  This is achieved if the 
following two independent conditions are satisfied
\begin{align}\label{eq:epsRelnu}
\epsilon^S_{ij}&\sim \epsilon^H_{ij}\,\epsilon^\Sigma_{ij}\,,&\epsilon^\Phi_{ij}&\sim(\epsilon^S_{ij})^2~.
\end{align}
Employing these two conditions, the parametric expression of the neutrino mass eigenvalues is
\begin{align}
\mnui  \sim \left( \epsilon^H_{12}  \epsilon^H_{23} \right)^2 \frac{ \langle H_3\rangle^2 }{ \langle \Sigma_1\rangle^2} \langle\Phi_1\rangle
~\to~
\frac{m_u^2}{m_t^2}  \frac{ v^2}{ \Lambda_{\rm UV} }~,
\label{eq:ISSPS3}
\end{align}
where the last term follows from the assumption of a unique scale for the VEVs of the fields dominantly 
localized on the first site ($\langle \Sigma_1\rangle \sim \langle\Phi_1\rangle \sim \Lambda_{\rm UV}$).  This expression 
is analogous to the standard type-I seesaw formula, with the extra factor  
$(\epsilon^H_{12}  \epsilon^H_{23})^2 \sim m_u^2/m^2_t  \sim 10^{-8}$, which is the (double) price to pay to connect the EWSB scale (localized on the third site) to the scale of lepton-number 
breaking (localized on the first site).

Setting $10^{-2} \lsim m_\nu \lsim 10^{-1}~{\rm eV}$, as suggested by cosmological observations 
and neutrino data, the relation in~(\ref{eq:ISSPS3}) implies $\Lambda_{\rm UV} \sim 10^4$~TeV,
providing a non-trivial consistency check of our construction. This is indeed the correct order of magnitude 
for the ${\rm PS}_1$ breaking scale, once we assume 
$\langle \Sigma_3 \rangle \sim 1$~TeV for a solution of the $B$-physics anomalies (see next section).

In summary, the phenomenological requirements of an anarchical neutrino mass matrix
and the solution of the $B$-physics anomalies, allow us to 
 fix completely all the scales of the problem. 
 Adjusting the $\cO(1)$ factors, we end up with the following eigenvalues 
 for $m_R$, $\mu$ and $m_D$:
\begin{align}\label{eq:muVEV}
\begin{aligned}
\mDi \sim \mui &\sim(10^{-2},\,1,\,10^2)~\mathrm{GeV}\,,  \\
\mRi &\sim(10^7,\,10^5,\,10^3)~\mathrm{GeV}\,,  \\
\mu_i &\sim(10^7,10^{-1},10^{-9})~\mathrm{GeV}\,.
\end{aligned}
\end{align}
Note that these scales hierarchies are for reference only, since their strong sensitivity to the $\mathcal{O}(1)$ numbers can make them vary up to an order of magnitude.

\subsection{PMNS unitarity violation}
The extended structure of the neutrino sector, involving the mixing of 9 independent neutrino states, 
 necessarily leads to unitarity violations in the Pontecorvo--Maki--Nakagawa--Sakata (PMNS)  matrix,
which controls  the mixing of the  light sub-sector only.  We describe this effect 
 defining the matrix  $\eta = |{\bf 1} - NN^{\dagger}|$, where $N$ is the non-unitary $3\times 3$ PMNS matrix. To leading order in the ISS expansion, the Hermitian matrix $\eta$ can be computed as (see Appendix~\ref{app:NeutrinoDetails} for details)
\begin{equation}
\eta \approx m_D\, m_R^{-1}\,(m_D\,m_R^{-1})^{\dagger}\,.
\end{equation}
Assuming Yukawa couplings of order one, the expected size of the PMNS unitarity violations at order $(\epsilon^{L}_{ij})^{4}$ is
\begin{equation}
|\eta| \sim \bigg| \frac{\mDt}{\mRt}\bigg|^{2} 
\begin{pmatrix}
(\epsilon_{12}^{L}\epsilon_{23}^{L} )^{2} & \epsilon_{12}^{L} (\epsilon_{23}^{L} )^{2} &\epsilon_{12}^{L}\epsilon_{23}^{L} \\
 \epsilon_{12}^{L} (\epsilon_{23}^{L} )^{2}  & (\epsilon_{23}^{L} )^{2} &\epsilon_{23}^{L} \\
\epsilon_{12}^{L}\epsilon_{23}^{L}& \epsilon_{23}^{L}  & 1
\end{pmatrix} \,.
\end{equation}
As can be seen,  
the largest violation occurs in $|\eta_{33}|$, i.e.~in the unitarity sum of the third row or third column of the PMNS matrix, 
as pointed out first in \cite{Greljo:2018tuh}. This effect propagates  to the other sites via appropriate insertions of $\epsilon^{L}_{ij}$, 
resulting in a texture for $\eta$ that naturally suppresses unitarity violations involving the light families. Given $\epsilon_{ij}^{L} \sim 0.1$ as set by quark mixing, the current bounds on unitarity violation~\cite{Antusch:2014woa,Antusch:2015mia} are automatically satisfied if $|\eta_{33}|$ obeys
\begin{equation}
|\eta_{33}| \approx \bigg| \frac{\mDt}{\mRt}\bigg|^{2} < 5.3\times 10^{-3}\,.
\end{equation}
For $\mRt \sim 10^{3}$ GeV, this bound is satisfied requiring $\mDt \lesssim 0.4 \, m_{t} \approx 70$ GeV. 
This $\cO(1)$ correction to the unbroken PS relation $\mDt = m_{t}$
is similar to the splitting occurring in the down sector, where $m_{\tau} \approx 0.8 \, m_{b}$ at 2 TeV.
We thus deduce no fine-tuning of the Dirac neutrino mass is required for the success of the mechanism.
On the other hand, the construction naturally leads to unitarity violations close to the present experimental  
bounds. 

\subsection{$U(1)_F$ breaking}
 \label{sect:AxiMajo}
In the absence of nearest-neighbor interactions, the model respects a global $U(1)_{F_1}\times U(1)_{F_2} \times U(1)_{F_3}$ fermion number symmetry. When these interactions are turned on, this symmetry is explicitly broken to the diagonal $U(1)_F$ subgroup. The VEV of $\Phi$,
which is localized mainly in the first site, spontaneously breaks $U(1)_F$ giving rise  to a massless Nambu-Goldstone (NG) boson. As is well known, baryon and lepton number emerge as accidental global symmetries when PS is broken to the SM by the VEV of $\Sigma$. From the structure of $\mathcal{L}_{\nu}$, it is clear that the spontaneous breaking of fermion number occurs only in the leptonic sector. Therefore, baryon number remains as an accidental global symmetry of the theory, ensuring proton stability, while we can identify 
the NG boson with the Majoron $J$. 

Since the breaking of $U(1)_F$ is localized on the first site, the Majoron is mostly the phase of $\Phi_{1}$ and therefore couples to $S_L^{\scriptscriptstyle (i)}$ as
\begin{equation}
\mathcal{L}_{J} \supset  \frac{i}{2}\frac{J}{\langle \Phi_{1} \rangle} \mu_{i} \, \bar S_{L}^{\scriptscriptstyle (i)}{S_{L}^{\scriptscriptstyle(i)}}^c \,.
\end{equation}
To give a mass to the Majoron, $U(1)_F$ must be explicitly broken.
On general grounds, 
one expects all global symmetries to be explicitly broken by Planck-scale
dynamics. Here we assume that such breaking takes place at a generic 
high scale $\Lambda_{F} \gg \langle \Phi_{1} \rangle$ via dimension-5 operators localized in the first site
\begin{equation}
\mathcal{L} _{J}\supset c_{1}\frac{\Phi_{1}^{5}}{\Lambda_{F}} + c_{2}\frac{\Phi_{1}^{4}\Phi_{1}^{*}}{\Lambda_{F}} + c_{3}\frac{\Phi_{1}^{3}\Phi_{1}^{*2}}{\Lambda_{F}} + {\rm h.c.} \,,
\end{equation}
resulting in a Majoron mass
\begin{align}
m_{J}^{2} \simeq \frac{1}{2}\left(25\,c_{1} + 9\,c_{2}+c_{3}\right)\frac{\langle \Phi_{1} \rangle^{3}}{\Lambda_{F}} \,.
\end{align}
If $U(1)_F$ is explicitly broken at the Planck scale (i.e.~for $\Lambda_{F} = M_P$), the resulting Majoron mass is around $50~\mathrm{GeV}$ for $\langle \Phi_{1} \rangle \sim 10^7$ GeV and $c_{i} \sim 1$. In this case, the Majoron decays to active neutrinos with a width controlled 
by\footnote{Potential decays to $S_L^{\scriptscriptstyle (3)},\nu_{R}^{\scriptscriptstyle (3)}$ are similarly suppressed by $(\mu_{3}/\langle \Phi_{1} \rangle)^{2}$.} 
 $(m_{\nu} / \langle \Phi_{1} \rangle)^{2}$. The 
 resulting Majoron is long lived on cosmological time scales, potentially leading to an unacceptably large relic abundance~\cite{Akhmedov:1992hi}. However, for $\Lambda_{F} \lesssim 10^{12}$~GeV, we have $m_{J} \gtrsim m_{R}^{\scriptscriptstyle (2)}$ and the Majoron decays promptly to the pseudo-Dirac pair $S_{L}^{\scriptscriptstyle (2)}, \nu_{R}^{\scriptscriptstyle (2)}$ with a lifetime of $\tau \sim 1$ ps. 
 
As a final point, we note that the would-be NG bosons in the coset $U(1)_{F_1}\times U(1)_{F_2} \times U(1)_{F_3}/ U(1)_F$ 
correspond to global symmetries with large explicit breakings $\propto \langle \Omega_{ij} \rangle$. These  would-be NG thus 
acquire masses of order $f_{ij}$ and have unsuppressed decays into the heavy pseudo-Dirac neutrinos.

\section{Towards a Continuous Higher-Dimensional UV completion}\label{sec:5D}
Up until now we have described the ${\rm PS}^{3}$ model using the discrete language of three sites connected by nearest-neighbor interactions.
This language implicitly admits the embedding of the theory into a higher-dimensional discrete construction~\cite{ArkaniHamed:2001ca,Hill:2000mu,Cheng:2001vd}:
 non-local interactions with respect to the additional, discrete, space-like coordinate are suppressed 
by appropriate scale ratios, whose net effect is encoded by the $\epsilon^{F}_{ij}$ spurions ($F = L, H,\ldots$).  

However, the underlying higher-dimensional completion of the model does not need not be discrete: the three sites can be viewed as special (discrete) 
values $y_{i}$ of a continuous, compact, space-like coordinate $y$. These special points indicate where 
the three generations of fermions are localized, or better quasi-localized, in the $y$ space.
In this context, the power-like behavior of the scale ratios  naturally leads us to promote the 
$\epsilon^{F}_{ij}$ to exponential functions of the continuous variable $y$:
\begin{equation}
\epsilon_{ij}^{F} \equiv \epsilon_{F} (y_i-y_j)  = e^{-M_{F} / f_{ij}  } \,, \quad   f_{ij}= |y_{i} - y_{j}|^{-1} \,.
\label{eq:expprof}
\end{equation}
Here $M_{F}$ is some mass scale unique to each field. In this description, the matter fields (scalar and fermions) can be thought of as having exponentially decaying {\em profiles} in $y$  space, where the steepness of the profile is controlled by $M_F$. With $\cO(1)$ differences in the $M_{F}$ values we easily achieve the large hierarchies outlined in~\eqref{eq:muVEV}. Moreover, with an almost equal spacing among the three notable points 
($f_{12} \sim  f_{23}$) we achieve the geometric behavior of the scale ratios implied by the flavor hierarchies.

As we discuss in Section~\ref{sec:naiveFlavor}, the assumption in~\eqref{eq:expprof} provides a transparent 
and robust interpretation of the whole flavor structure of the model, as arising 
from an extended spacetime characterized by coordinates $\{y,x_\mu\}$, with $y$ being continuous and compact.
Interestingly enough, this is achieved without the need 
of specifying in detail the geometry of the extra dimension: they only key requirement is a setup allowing
$y$ profiles as in (\ref{eq:expprof}) to be stable field configurations.  More precise hypotheses about the geometry are needed to 
address the gauge sector, which presents some problematic points as discussed in Section~\ref{sect:naivegauge}. 
The sketch of a complete dynamical 5D model is presented in Section~\ref{sect:full5D}.

\subsection{Flavor sector}
\label{sec:naiveFlavor}
We can infer the localization properties of the matter fields in this space by matching their profiles onto the corresponding $\epsilon^{F}_{ij}$ spurions of the discrete picture. In particular, the absence of $\epsilon_{ij}^R$ in the discrete model implies that right-handed fermions should be highly localized at each site. For now, we make the fully localized approximation $\Psi_{R}^{\scriptscriptstyle (i)} (y) \propto \delta (y-y_{i})$, corresponding to the limit $M_{R}/ M_{L} \to \infty$. However, 
$\Psi_{R}$ can be somewhat delocalized as we quantify later.  On the other hand, in order to generate non-vanishing $\epsilon_{ij}^{L}$  (i.e.~non-vanishing 
left-handed mixing) we assume the left-handed fermions to be rather delocalized, with profiles behaving as
\begin{equation}
\Psi_{L}^{\scriptscriptstyle (i)} (y) \propto e^{-M_{L} |y-y_{i}|}  \,,
\end{equation}
and similarly for $S_L^{\scriptscriptstyle (i)}$. 

The scalar fields must be localized such that their VEVs give the largest symmetry breaking in the appropriate family, as discussed in detail in Sections~\ref{sec:PS3rev} and~\ref{sec:nu}. Specifically, as electroweak symmetry breaking is largest in the third family, we take the Higgs VEV to be localized at $y_{3}$
\begin{equation}
\langle H  (y) \rangle \sim \langle H_{3} \rangle  \, e^{-M_{H} | y- y_{3} |} \,.
\end{equation} 
In contrast, $SU(4)$ and $U(1)_{F}$ breakings are largest in the first family, suggesting localization of the $\Sigma$ and $\Phi$ VEVs at $y_{1}$ 
\begin{align}
\begin{aligned}
\langle \Sigma (y)  \rangle &\sim \langle \Sigma_{1} \rangle  \, e^{-M_{\Sigma} | y- y_{1} |} \,,  \\
\langle \Phi  (y) \rangle &\sim \langle \Phi_{1} \rangle  \, e^{-M_{\Phi} | y- y_{1} |} \,.
\end{aligned}
\end{align} 
In this continuous construction, the coupling structure for fields depending only on $x_{\mu}$ is given by integrating over $y$, giving the overlap of the field profiles. For instance, the Dirac mass matrix is computed as
\begin{align}
& m_{D}^{ij}  \left(  \bar{\Psi}_{L}^{\scriptscriptstyle (i)}   \Psi_{R}^{\scriptscriptstyle (j)}  \right)_{\rm 4D}
 \propto \int dy \, \bar{\Psi}_{L}^{\scriptscriptstyle (i)}  (y) \langle  H   (y) \rangle \, \Psi_{R}^{\scriptscriptstyle (j)}  (y) \nonumber \\
& \longrightarrow \,  m_{D}^{ij}  \sim  \langle H_{3} \rangle \, e^{-M_{H} |y_{j}-y_{3}|} e^{-M_{L} |y_{i}-y_{j}|}  \,,
\label{eq:mDcont}
\end{align}
which reproduces the structure of~\eqref{eq:massMats} up to a chirally-suppressed right-handed rotation that we neglect. This also gives us a way to quantify the degree of localization of $\Psi_{R}$, as delocalizations of size $\epsilon^{R}_{ij} \lesssim \epsilon_{ij}^{L} \epsilon_{ij}^{H}$ yield right-handed rotations smaller than those we neglect. This explanation of the SM Yukawa hierarchies is similar to the ones proposed in~\cite{Dvali:2000ha,Panico:2016ull}.

\begin{figure}[t]
\centering
\includegraphics[scale=0.135]{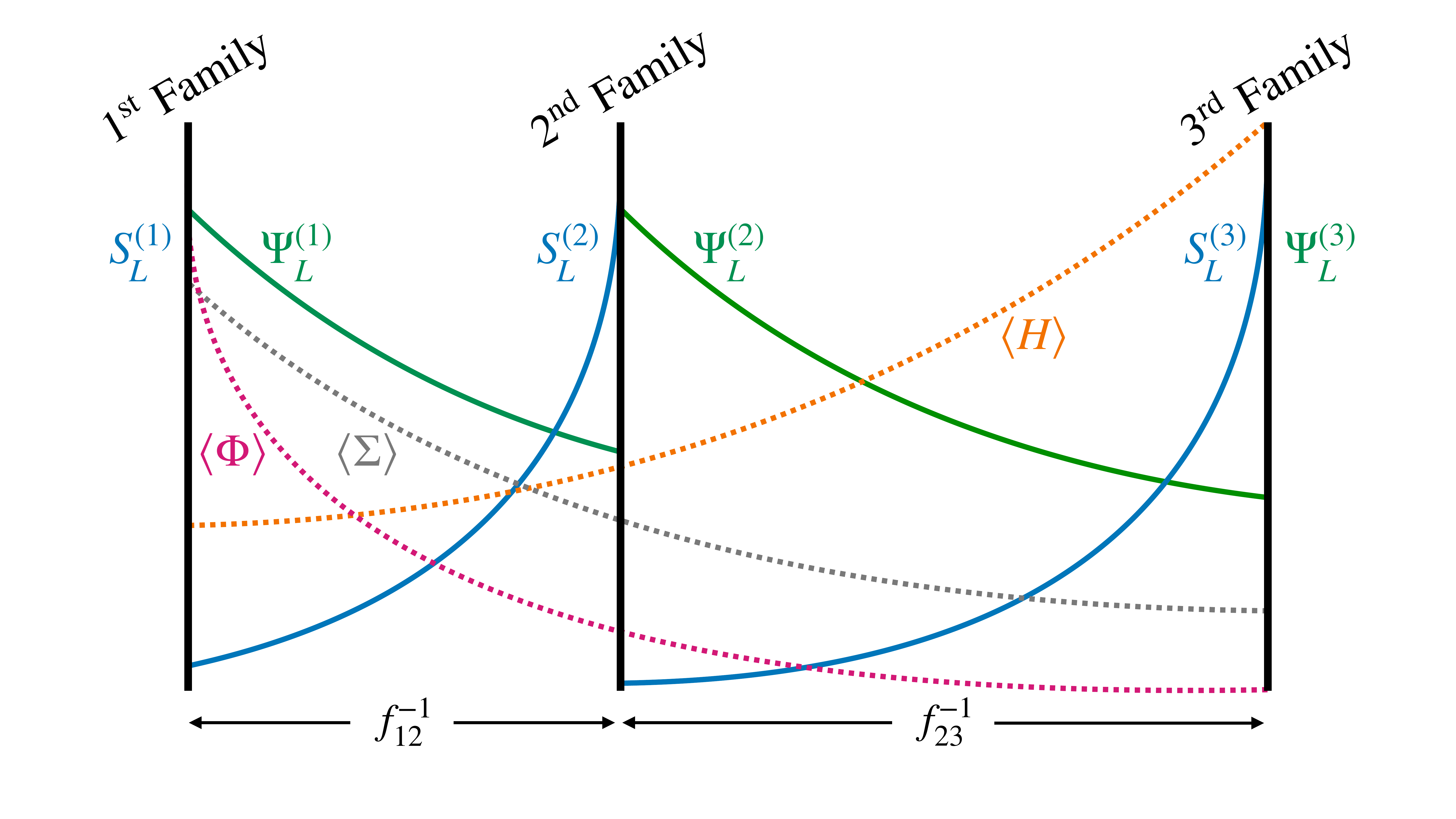}
\caption{Schematic representation of the field profiles in the continuous 5D  version of the model. 
Right-handed fields, being mostly localized at their respective sites, are not shown. Not shown are also the 
parts of fermion profiles that can be rotated away by a suitable basis choice or that generate chirally-suppressed rotations.}
\label{fig:profiles}
\end{figure} 

Comparing~\eqref{eq:massMats} and~\eqref{eq:mDcont}, we can 
deduce the structure for the  $\epsilon_{ij}^{F}$ anticipated in  (\ref{eq:expprof}),
showing the emergence of the discrete picture spurions from the continuous one. 
The continuous setup here can be qualitatively summarized by the schematic plot in Fig.~\ref{fig:profiles} where, for additional clarity, 
we do not show the parts of the fermion profiles which generate rotations that are chirally suppressed. 
Note that, at this level of the description, we do not need to specify if the different flavors of a given 
fermion field are generated by a unique 5D field, with a multi-cusp profile, or if they arise from three independent fields with a single-cusp profile.
We will  come back to this point in Section~\ref{sect:full5D}. 

The mass parameters and distances in the continuous construction can be deduced from the size of $\epsilon_{ij}^{F}$. In particular, we have
\begin{equation}
-\frac{M_{F}}{f_{ij} } \approx \log \epsilon_{ij}^{F} \,,
\end{equation}
which also fixes the ratio of interfamily distances
\begin{equation}
\frac{f_{23}}{f_{12}} \approx \frac{ \log \epsilon_{12}^{F}}{ \log \epsilon_{23}^{F}} \approx \frac{2}{3} \,,
\label{eq:fratios}
\end{equation}
where we have used $\epsilon_{12}^{L} \approx 2\epsilon_{23}^{L} \approx 0.2$ as pointed out in~\eqref{eq:epsLandH}. 
As discussed in the earlier sections, achieving successful fermion masses and mixing requires non-trivial relations between the $\epsilon_{ij}^{F}$, c.f.~\eqref{eq:epsLandH} and~\eqref{eq:epsRelnu}. In the continuous higher-dimensional picture, these translate into a series of rather simple 5D mass relations:\footnote{Also implied is the inequality $M_{\Phi} > M_{S} > M_{\Sigma}  > M_{H} > M_{L}$, which tells us that $\Phi$ has the steepest (most localized) profile while $\Psi_{L}$ is the most spread. The parameter $M_{R}$, characterizing the steepness of the right-handed profiles, should satisfy $M_{R} \gtrsim 3M_{L}$.}
\begin{equation}
M_{H} \approx 2M_{L}\,, \hspace{3.5mm} M_{\Phi} \approx 2M_{S} \,, \hspace{3.5mm} M_{\Sigma} \approx M_{S} - 2M_{L} \,.
\label{eq:bulkM}
\end{equation}
Due to these relations, all $\epsilon^F_{ij}$ spurions are fixed in terms of only two fermion mass parameters, $M_L$ and $M_S$, given in units of the basic scales $f_{ij}$. Note however
that the flavor structure of the model 
does not depend on the overall 5D scale
\begin{equation}
L = f_{12}^{-1} + f_{23}^{-1}~.
\label{eq:L5D}
\end{equation}

\subsection{Gauge sector}
\label{sect:naivegauge}
In the discrete picture the masses of the gauge bosons are $m_{ij}^{\cG} \sim  g_\cG f_{ij}^{\cG}$  where,
following the notation of Section~\ref{sec:PS3rev},   $f_{ij}^{\cG}$ denotes the VEV of the 4D link field  $\Omega_{ij}^{\cG}$
in (\ref{eq:linkfields}) associated to the group $\cG = 4,L,R$.
Following the discussion in Sec.~\ref{sec:naiveFlavor}, it is tempting to transform the discrete set of 4D link fields 
into a continuous set of link fields, 
\begin{align}
\Omega_{ij}^{\mathcal{G}}(x_{\mu}) &\rightarrow  \, \Omega(x_{\mu}, y) \,, 
\end{align}
identifying the $f_{ij}^{\cG}$ with the ${\cG}$-independent $f_{ij}$ defined in 
(\ref{eq:expprof}) and characterizing the overall 5D scale via~\eqref{eq:L5D}.
However, as we shall see, this introduces a series of problems.

On general grounds, when a (5D) continuous link field takes a VEV, an infinite number of PS gauge groups are horizontally Higgsed down to the diagonal subgroup. 
The finiteness of the space sets an infrared cutoff $m_{\rm IR} \sim 4\pi L^{-1}$ and quantizes the mass of any excitations in units of 
this scale, at least in a flat geometry. Thus, we end up with one set of massless gauge bosons transforming in the adjoint representation of the diagonal PS group,\footnote{The diagonal PS group is vertically broken to the SM by the VEV of $\Sigma$, so the coset PS/SM contains only massive gauge bosons.}
 plus a numerable infinite tower of increasingly heavier gauge bosons  with masses $m_{V_n} \sim n / L$ (again in a flat geometry). 
 
Within this context, the discrete link field VEV structure in Section~\ref{sec:PS3rev} is recovered when
considering only the lightest excitations associated to the  two distances among the the three special points 
($f_{12}^{-1}$ and $f_{23}^{-1}$), for which 
\begin{equation}
\langle \Omega_{ij} \rangle \sim n_{ij} / L\,,
\end{equation}
with $n_{ij} \sim L f_{ij}$. If one hopes to address the $B$-anomalies with the lightest $SU(4)$ excitations, the scale  
$f_{23}$ should be in the TeV range. This poses two problems
\begin{itemize}
\item[(I)]
In the na\"ive picture illustrated above, all the links associated to a given length have the same VEV.
Allowing for $\langle\Omega_{ij}^{4} \rangle < \langle\Omega_{ij}^{L,R} \rangle$, as required by the tight constraints from $Z$ and $W$ coupling modifications,
seems to suggest different {\em labels} on the continuous coordinates associate to the different gauge groups,
i.e.~$\Omega_{ij}^{\mathcal{G}}(x_{\mu}) \rightarrow  \, \Omega(x_{\mu}, y_\cG)$.

\item[(II)] The successful reproduction of fermion masses and mixings presented 
in Section~\ref{sec:naiveFlavor} indicates $f_{12} \sim f_{23}$, which would imply that that also 
$f_{12}$ should be in the TeV range, in manifest conflict with the FCNC bounds.  
Alternatively, $L^{-1}$ could be pushed above $10^{3}$ TeV, which is still compatible with a good description of the 
spectrum but precludes a solution of the  $B$-anomalies.
\end{itemize}
There are different ways to address these problems depending on the specific geometry of the extra-dimensional
construction, as well as on additional details of the UV structure of the model. For instance, concerning problem I, 
$\cO(1)$ differences among the $\langle \Omega_{ij}^{\cG} \rangle$ belongings to different gauge groups (on the same interval) 
could effectively be obtained adding additional source of breaking and/or placing non-universal kinetic terms 
for the different gauge groups on the three 4D branes.
The problem II, which is the most serious one, requires a parametric decoupling of the IR cutoff controlling the masses 
of the vector resonances from the inverse size of the interval. As we discuss in detail below, this effect is naturally 
achieved in the context of a warped geometry.

\subsection{Warped 5D Model}
\label{sect:full5D}
A natural and more precise interpretation of the ideas sketched above can be obtained 
in terms of a physical fifth dimension compactified on an interval $y \in [0,L]$ containing 3 four-dimensional branes or topological defects.\footnote{These defects could be physical branes, domain walls supporting fermion localization, or a combination of both~\cite{Rubakov:1983bb,Jackiw:1975fn,Weinberg:1981eu,Kaplan:1995pe,ArkaniHamed:1999dc}.}
For simplicity, we take the interval to be bounded by two of these branes.  An essential property of these defects is that they allow for the localization of fermions, while the endpoint branes additionally impose boundary conditions for all fields. The gauge group in the 5D bulk is taken to be a single instance of PS, namely $ {\rm PS_{5D}} = SU(4) \times SU(2)_L \times SU(2)_R$, where the 5D matter content is charged under ${\rm PS_{5D}}$ as indicated in Table~\ref{tab:5Dmat}.

\begin{table}[t]
\begin{center}
\begin{tabular}{|c|ccc||c|}
\hline
Fields & $SU(4)$ & $SU(2)_{L}$ & $SU(2)_{R}$ & $U(1)_F$  \\
\hline
\hline
$\Psi_{L}$ & $\mathbf{4}$ & $\mathbf{2}$ & $\mathbf{1}$ & $1$ \\  
$\Psi_{R}$ & $\mathbf{4}$ & $\mathbf{1}$ & $\mathbf{2}$ & $1$ \\  
$S_{L}$ & $\mathbf{1}$ & $\mathbf{1}$ & $\mathbf{1}$ & $1$ \\  
\hline
\hline
$\Sigma$ & $\mathbf{4}$ & $\mathbf{1}$ & $\mathbf{2}$ & $0$ \\  
$H$ & $\mathbf{1}$ & $\mathbf{2}$ & $\mathbf{\bar 2}$ & 0 \\  
$\Phi$ & $\mathbf{1}$ & $\mathbf{1}$ & $\mathbf{1}$ & $2$ \\  
\hline
\end{tabular}
\caption{5D matter content with charges under $ {\rm PS_{5D}} = SU(4) \times SU(2)_L \times SU(2)_R$ as well as under the fermion number global symmetry $U(1)_{F}$.
Here, any generation indices are left implicit. \label{tab:5Dmat}}
\end{center}
\end{table}

 As anticipated, to address problem II we consider a warped spacetime geometry
\begin{equation}
ds^{2} = e^{-2\sigma(y)} \eta_{\mu\nu} dx^{\mu} dx^{\nu} -dy^{2} \,,
\label{eq:metric}
\end{equation}
where $\eta_{\mu\nu} = {\rm diag}(1,-1,-1,-1)$ is the 4D metric and the warp factor $\sigma$ is a monotonically increasing function of $y$ satisfying $\sigma(0) = 0$. The case where $\sigma(y) = ky$ corresponds to 5D anti-de Sitter (AdS$_{5}$) spacetime with constant curvature $k$. While we focus on AdS$_{5}$ in what follows, our results are easily generalizable to other warp factors. The relevant terms in the 5D action can be written as\footnote{We do not attempt to provide a dynamical origin for the fermion masses, as in~\cite{Ahmed:2019zxm,Carmona:2020uqx}, and simply assume that $M_i$ are $\mathbb{Z}_2$-odd parameters under the orbifold symmetry.}
\begin{align}
\begin{aligned}
&\mathcal{S}_{\rm 5D}  \supset  -\int d^5 x \sqrt{g} \, \Big[ M_L \bar \Psi_{L}\Psi_{L} +M_R \bar\Psi_{R} \Psi_{R}   \\
& + y\, \bar \Psi_{L}H\,\Psi_{R} + \tilde{y}\,  \bar \Psi_{L}\widetilde{H}\,\Psi_{R} + y_{\Sigma}\, \bar{S}_{L} \, \Sigma^{\dagger} \Psi_{R}  \\
& +  M_{S} \, \bar{S}_{L} S_{L} + y_{\Phi} \,\bar{S}_{L} \,\Phi  \, S_{L}^{c} + V(H,\Sigma,\Phi)  + \, {\rm h.c.} \Big]\,,
\end{aligned}
\end{align}
where $V(H,\Sigma,\Phi)$ is the most general potential of the scalar fields and any generation indices have been 
left implicit.\footnote{The three families of fermions could arise directly from three 5D fields, or perhaps as multiple zero modes of the same 5D field as in Ref.~\cite{Dvali:2000ha}.}
In order to study the low energy 4D dynamics of this theory, we perform a KK-decomposition of each 5D field $X_{\rm 5D}$ as
\begin{equation}
X_{\rm 5D} (x^{\mu} ,y) = \frac{e^{sky}}{\sqrt{L}} \sum_{n=0}^{\infty} f_{n}(y) X_{n}(x^{\mu}) \,,
\label{eq:KKdecomp}
\end{equation}
where $s=0,1,3/2$ for gauge bosons, scalars, fermions, in order to ensure the fields are canonically normalized.  The 4D fields $X_{n}(x^{\mu})$ in~\eqref{eq:KKdecomp} are the zero mode ($n=0$) and KK modes whose localization along the extra dimension are described by the orthonormal mode profiles $f_{n}(y)$.\footnote{The functions $f_{n}(y)$ satisfy the orthonormalization condition $L^{-1} \int_{0}^{L} dy\, f_{n}(y) f_m(y) = \delta_{nm}$.} Inserting this expansion into the action, we see that this geometry has the property that mass scales become exponentially suppressed as $Me^{-ky}$ for $y>0$. This leads to the identification of the $y=0$ ($y=L$) brane as the UV (IR) brane. In particular, the IR mass scale is $Me^{-kL}$, and thus $kL \gg 1$ corresponds to strong warping. We recall that a warp factor $kL \approx 37$ was proposed in~\cite{Randall:1999ee} as a way to address the hierarchy between the Planck and EW scales. Our aim here is to address a smaller hierarchy, namely the one between  $\Lambda_{\rm UV} \sim k \sim 10^4$~TeV and the EW scale, which still requires strong warping with $kL \approx 9$.

The couplings of the 4D fields $X_{n}(x^{\mu})$ are determined by the overlap of the KK mode profiles as e.g.
\begin{equation}
g_{{\rm 5D}} \int d^5 x \sqrt{g} \,\bar{\Psi} \slashed{A} \Psi = \frac{g_{\rm 5D}}{\sqrt{L}} \, \lambda_{lnm} \int d^4 x  \,\bar{\psi}_{n} \slashed{A}_{l} \psi_{m} \,,
\label{eq:gff}
\end{equation}
with
\begin{equation}
\lambda_{lnm} = \frac{1}{L} \int_{0}^{L} dy\, f_{n}^{\Psi}(y) f^{A}_{l}(y) f_{m}^{\Psi}(y) \,.
\end{equation}
This model reproduces the essential features of the flavor sector discussed in Section~\ref{sec:naiveFlavor}. The general features of warped extra dimensions are summarized in~\cite{Zyla:2020zbs}. In particular, the fermionic zero modes are made chiral by the usual orbifolding and have exponential profiles of the form
 \begin{equation}
f^{\Psi_{L,R}}_{0} = N_{L,R} \exp\left[ \left(\frac{1}{2}k \mp M_{L,R}\right) y \right] \,,
\end{equation}
where $N_{L,R}$ is a normalization factor. Thus, we can localize the fermions according to Fig.~\ref{fig:profiles} via a suitable choice of bulk masses. Similarly, the desired exponential scalar VEV profiles in the 5D bulk can be obtained via a suitable choice of boundary potentials which trigger SSB on the appropriate branes~\cite{Goldberger:1999uk,Cabrer:2010si,Cabrer:2011fb,Cabrer:2011vu,Quiros:2013yaa}. We now turn our attention to the gauge sector of this model, which requires a more detailed examination.

\subsubsection{Gauge sector}
The gauge boson mode profiles $f_{n}^{A}(y)$ satisfy
\begin{equation}
\partial_{y}^{2} f_{n}^{A} - 2k \,\partial_{y} f_{n}^{A} + m_{n}^{2} e^{2ky} f_{n}^{A} = 0 \,,
\label{eq:gaugeEoM}
\end{equation}
where the KK masses $m_{n}$ are dominantly determined by the IR boundary condition (BC). 
We are always interested in the case where the IR boundary respects the bulk gauge symmetry. This is achieved imposing 
the Neumann condition $\partial_{y} f_{n}^{A} |_{y=L} = 0$, which leads to the following expression for the KK masses~\cite{Pomarol:1999ad}
\begin{equation}
\label{eq:KKmass}
m_{n} \approx \left(n-\frac{1}{4} \right) \pi k \,  e^{-kL}~, \qquad (n \geq 1)~.
\end{equation}
As expected, these masses are {\em warped down} from the UV scale $k$. 
If the UV boundary also respects ${\rm PS_{5D}}$, we have $\partial_{y} f_{n}^{A} |_{y=0,L} = 0$, and there is also a massless zero mode with a flat profile in $y$ (corresponding to an unbroken 4D gauge symmetry). 
This universal zero mode can be seen immediately as the solution of~\eqref{eq:gaugeEoM} with $m_{0} = 0$, which yields the normalized mode $f^{A}_{0}(y) = 1$.
From gauge vertices involving the zero mode $A_{\mu,0} (x^\mu)$, such as~\eqref{eq:gff} (where $\lambda_{0nm} = \delta_{nm}$), one can make the tree-level identification $g_{\rm 4D} = g_{\rm 5D} / \sqrt{L}$. 

For $m_{n} \neq 0$,~\eqref{eq:gaugeEoM} 
has a general solution in terms of Bessel functions. A reasonable approximation of the first KK mode profile which captures all the important physics is
\begin{align}
f_{1}^{A}(y) \approx \sqrt{kL} \left(e^{2k(y-L)} - c_{\rm UV} \right) \,,
\end{align}
where $c_{\rm UV}$ is a constant fixed by the UV boundary condition for each gauge generator. 

The bulk ${\rm PS_{5D}}$ gauge group is vertically broken to the SM by the VEV of $\Sigma$ dominantly localized on the first family brane
(i.e.~at $y=0$). This removes the zero modes in the coset PS/SM. Because of the exponentially falling nature of $\langle \Sigma \rangle $, this breaking can be treated via the UV BC, which can be approximated by the Dirichlet condition 
$f_{n}^{A} |_{y=0} \approx 0$ in the limit $\langle \Sigma_1 \rangle \gtrsim k$.\footnote{The fully correct procedure would be to include a bulk mass $M_{A}(y)$ for the gauge field in~\eqref{eq:gaugeEoM}, which yields a similar result when $\langle \Sigma \rangle $ is exponentially localized in the UV.}
In contrast, the SM generators remain unbroken and thus have Neumann conditions both in the UV and in the IR. 
The value of $c_{\rm UV}$ in each case is approximately given by
\begin{equation}
c_{\rm UV} ({\rm D}) \sim e^{-2kL} \,,\qquad c_{\rm UV} ({\rm N}) \sim \frac{1}{kL}\,,
\end{equation}
where N (D) indicates Neumann (Dirichlet) BCs at the UV boundary. 

In both cases, the KK modes have flavor non-universal profiles peaked in the IR (i.e.~at $y=L$), where the third family fermions are localized. All KK states thus have enhanced 4D couplings to the third family of size $g_{\rm KK}^{\scriptscriptstyle(3)} \approx g_{\rm 4D} \sqrt{kL}$. Away from the IR boundary, the couplings of the KK modes belonging to the PS/SM coset (which includes the $U_1$ LQ) become exponentially suppressed. As a result, to a good approximation these KK modes couple only to the third family. On the other hand, the KK mode profiles of the SM gauge group exponentially approach a constant value, leading to nearly universal light-family couplings of size $g_{\rm KK}^{\scriptscriptstyle(1)} \approx g_{\rm KK}^{\scriptscriptstyle(2)} \approx -g_{\rm 4D} /\sqrt{kL}$. In the limit of fully localized fermions, the deviation from universality in the light families is exponentially suppressed as $e^{-2k/f_{32}}$.

It is clear that the 4D description of this theory differs from that of the 4D (or 5D discrete) ${\rm PS}^3$ model, due to the appearance of an infinite tower of KK states. However, an interesting comparison is obtained considering 
the first KK modes of this construction and the massive gauge bosons in the (4D) non-universal 
4321 models
(see~\cite{Greljo:2018tuh,DiLuzio:2018zxy,Cornella:2019hct,Fuentes-Martin:2019ign,Fuentes-Martin:2020luw,Fuentes-Martin:2020hvc}).
The latter are indeed the the low-energy limit of ${\rm PS}^3$, as-well other third-family PS unification 
models~\cite{Fuentes-Martin:2020bnh}.  As we have seen, the KK modes of unbroken 4D gauge symmetries 
couple to IR (UV) localized states with a volume enhancement (suppression) of $\sqrt{kL}$.  This is for instance 
the case of the first KK mode of $SU(3)_c$, that we can identify with the coloron in 4321. 
In this case, the zero mode is the SM gluon, which must couple with strength $g_{\rm 4D} = g_{s}$. As a result, the first KK of 
$SU(3)_c$ couples to the third family with strength $g_{s} \sqrt{kL}$, that we can identify with 
the $SU(4)$ gauge coupling ($g_4$) in the 4321 model. 
In the limit of fully localized fermions, the couplings of this massive vector to the different fermion families are then
\begin{align}
g_{s}\sqrt{kL} \left(-\frac{1}{kL},-\frac{1}{kL},1 \right) = g_{4} \left(-\frac{g_{s}^{2}}{g_{4}^{2}},-\frac{g_{s}^{2}}{g_{4}^{2}},1 \right) \,,
\end{align} 
up to the previously mentioned exponentially suppressed deviation from universality in the light families. 
These are precisely the coloron couplings in the 4321 model. What is particularly remarkable in the 5D construction 
is the emergence of the $U(2)^5$ flavor symmetry as result of the (approximate) flatness of the KK profiles in the UV. 
A similar coupling structure holds for the first KK mode of SM hypercharge, that we can identify 
with the $Z^\prime$ in 4321. 

On the other hand, the LQ is a KK mode in the same multiplet as the coloron, but it corresponds 
to a symmetry broken in the UV and therefore has exponentially small couplings to the light families (again as in the 4321 model, before fermion mass mixing). With $kL\approx 9$, we have the LQ coupling $g_{4} \approx 3$, which is known to give a good fit to the flavor anomalies. Interestingly, a warp factor $kL \approx 9$ is also simultaneously compatible with sufficient FCNC suppression in right-handed currents $e^{-2kf_{23}} \sim 10^{-5}$ and realizing the correct fermion mass hierarchy without fine-tuning. For a Higgs VEV profile of the form $\langle H\rangle \propto e^{aky}$, the correct fermion mass hierarchy is achieved for $a\approx2$, which is the maximally spread Higgs VEV profile compatible with stabilizing the hierarchy between $\Lambda_{\rm UV}$ and the TeV scale~\cite{Cabrer:2010si,Cabrer:2011fb,Cabrer:2011vu,Quiros:2013yaa,Carmona:2011ib}.

What is not present in the 4321 model are the states corresponding to the KK modes of $SU(2)_L \times SU(2)_{R}$.
In the original (4D) formulation of ${\rm PS}^3$ these states are pushed in the 10~TeV domain playing with the VEVs 
of the corresponding link fields. This is not possible in the continuous 5D construction. Here the  
KK mass scale is universal, as shown in~\eqref{eq:KKmass}. In order to address the flavor anomalies
we need 
\begin{equation}
m_{1} \approx  m_{\rm LQ} \lsim~5~{\rm TeV}~, 
\label{eq:MLQbound}
\end{equation}
where the last inequality follows from the requirement of a perturbative description 
(i.e.~$g_{4} \lsim 3$).

The  $SU(4)\times SU(2)_{R} / U(1)_Y$ KK modes are not problematic since there is no corresponding zero mode: in particular, 
the massive $W_R$ and $Z''$ states have profiles similar to the LQ,
with vanishing couplings to the light families and a smaller coupling to the third
generation, and do not mix with the SM gauge bosons. 
In contrast, the electroweak KK modes have profiles similar to that of the coloron, as they are KK modes of an unbroken 4D gauge symmetry. 
The coupling to third family fermions is fixed to be $g^{\scriptscriptstyle(3)}_{\rm KK} = g_{L} \sqrt{kL}$, where $g_{L}$ is the $SU(2)_L$ 4D coupling.
These KK modes mix with the  SM EW gauge bosons leading to potentially dangerous volume-enhanced modifications of their 
couplings to third-generation fermions of the type
\begin{equation}
\frac{\delta g_{Z{\bar\psi_3}\psi_3}}{g_{Z\bar\psi_3\psi_3}} \sim \frac{m_{Z}^{2}}{m_{\rm LQ}^{2}} kL \,.
\label{eq:dgOng}
\end{equation}
These non-SM effects are highly constrained by data.
The strongest bound comes from $Z\rightarrow \tau_{L}\tau_{L}$, which sets a constraint on (\ref{eq:dgOng}) 
at the per-mil level~\cite{Efrati:2015eaa}. The $Z\rightarrow \tau_{L}\tau_{L}$ correction can be suppressed by an enlarged custodial symmetry, 
as in~\cite{Agashe:2006at}, but this cannot simultaneously protect the $W$ couplings, which also set a per-mil constraint 
from $\tau$ decays~\cite{Lusiani:2018zvr}.  Na\"ively, in order to safely satisfy these bounds we would need 
$m_{\rm LQ} \gtrsim 10$~TeV, which is in conflict with the requirement~(\ref{eq:MLQbound}).\footnote{The non-vanishing (SM-like) couplings to light states also imply 
relevant bounds from direct searches, especially on the $Z^\prime$. However, these are evaded for $m_{Z^\prime} \gsim 5$~TeV~\cite{CMS:2019tbu}.
Corrections to the oblique $S$ and $T$ parameters are under control due to the custodial $SU(2)_L \times SU(2)_R$ symmetry in the bulk, which is respected by the IR boundary~\cite{Agashe:2003zs}.} 

The tension between~\eqref{eq:MLQbound} and \eqref{eq:dgOng} arises because all of the KK modes share a common mass scale. This occurs for two reasons: 
 i) the IR boundary treats all generators of PS$_{\rm 5D}$ in the same way, and ii) there is only one fundamental length scale ($kL$). This suggests two directions to pursue 
in order to effectively decouple what we denoted as horizontal breaking of $SU(4)$ and $SU(2)_L$ in Section~\ref{sec:PS3rev}.
The first direction  is that of imposing different BCs for the generators in the coset PS$_{\rm 5D}$/SM. 
In particular, this can be achieved adding brane-localized kinetic terms with different weight for $SU(4)$ and $SU(2)_L$.
The second direction is that of considering a 6D model, separating the fundamental length scales for $SU(4)$ and $SU(2)_L \times SU(2)_R$. The topology of the 6D construction should be such the full PS group propagates in the space shared by the fermions.
Since the conflict between~\eqref{eq:MLQbound} and~\eqref{eq:dgOng} is only an $\mathcal{O}(1)$ factor, both these directions
are likely address this issue. Their detailed exploration is beyond the scope of this paper and will be presented elsewhere.  Here 
we simply note that  the conclusions regarding the neutrino and flavor constructions of Sects.~\ref{sec:nu} and \ref{sec:naiveFlavor} 
are expected to remain unchanged.

As a final note, the AdS/CFT correspondence~\cite{Maldacena:1997re,Witten:1998qj,Gubser:1998bc} provides a connection between our model in AdS$_{5}$ and a strongly coupled 4D conformal field theory (CFT) with a large number of colors $N \sim (4\pi/g_{4})^{2}$. The 4D CFT has a PS global symmetry where the SM subgroup is weakly gauged. Fields localized in the UV are elementary while IR localized fields, such as the LQ, correspond to composite resonances of the CFT. This is similar to the model proposed in Refs.~\cite{Barbieri:2016las,Barbieri:2017tuq},\footnote{This model has an extended global symmetry in order to realize the Higgs as pseudo-NG boson of the spontaneously broken CFT.} 
but with a specific flavor structure given by the three-site fermion localization in Fig.~\ref{fig:profiles}, which generates fermion mass hierarchies through a hierarchy of scales. The latter structure is, in its main features, in close correspondence with 
the 4D setup proposed in Ref.~\cite{Panico:2016ull}, which represents a change of paradigm compared to the widely discussed composite frameworks  where all the SM Yukawa operators are generated at the IR scale~\cite{Kaplan:1991dc,Grossman:1999ra,Gherghetta:2000qt,Huber:2000ie,Huber:2003tu}.

\section{Conclusions}\label{sec:conclusions}
  
The three-site Pati-Salam model~\cite{Bordone:2017bld} originates from the ambitious attempt to i)~unify and quantize 
the $U(1)$ charges of quark and leptons, ii)~obtain a natural description of all the SM Yukawa couplings  
in terms of $\mathcal{O}(1)$ parameters and fundamental scale ratios, and  iii)~address the recent hints of 
lepton-flavor non-universality violations in semileptonic $B$ decays. 
 
In this paper, we revisited this model with the twofold purpose 
of addressing the problem of neutrino masses as well as analyzing the possible embedding of this 
four-dimensional construction into an extra-dimensional model. These two issues, which at first sight appear to be rather 
disconnected, turn out to be closely related: contrary to the Yukawa couplings, whose hierarchical 
structure is largely insensitive to the UV completion of the theory, 
the anarchic structure of the neutrino mass matrix arises by 
the non-trivial interplay of low and high scales. The latter, which may appear {\em ad hoc} 
from a genuine 4D perspective, finds a natural explanation and can be better appreciated 
within the extra-dimensional embedding of the model.

The key feature of the model is the localization of electroweak symmetry breaking and the 
 breaking of the PS group at opposite sites: 
${\rm PS} \to {\rm SM}$ breaking 
 occurs on the first-generation (UV) site, while EWSB occurs
on the third-generation (IR) site. This breaking is 
communicated to the other sites via nearest-neighbor interactions
 suppressed by appropriate scale ratios which, in turn, are responsible 
for the hierarchical structure of the SM Yukawa couplings. 
We have shown that a realistic neutrino mass spectrum can be 
achieved with a minimal extension of the model featuring 
 three gauge-singlet fermions, with 
$U(1)_F$-breaking Majorana masses and $U(1)_F$-conserving 
couplings to the right-handed PS neutrinos. 
The highly hierarchical Majorana masses compensate the Dirac-type mass ratios via an effective site-by-site inverse seesaw mechanism, 
 leading to an anarchic light neutrino mass matrix.
 A non-trivial outcome of this construction is a prediction of neutrino 
 masses in the range $10^{-2} \lsim m_\nu \lsim 10^{-1}~{\rm eV}$,
under the assumption of a unique UV scale ($\Lambda_{\rm UV} \sim 10^4$~TeV)
controlling the ${\rm PS} \to {\rm SM}$  and $U(1)_F$ breaking.
This value of $\Lambda_{\rm UV}$ follows from a consistent description of the SM 
flavor hierarchies and coherent solution of the $B$-physics anomalies~\cite{Bordone:2017bld}. 
The construction also predicts one family of right-handed neutrinos at the TeV scale 
as well as sizable PMNS unitarity violations in the third family 
controlled by $ (m_t  / m_{R}^{\scriptscriptstyle (3)})^{2} \gsim 10^{-3}$, as 
noted first in~\cite{Greljo:2018tuh}.

The peculiar energy scales of the model
 find a natural interpretation in the extra-dimensional embedding, 
 where the three sites can be viewed as the special positions along the extra dimension where the fermion fields are quasi-localized.
In this context, the various scale ratios can naturally be understood as originating from scalar fields taking exponentially decaying VEVs in the bulk. 
In particular, the hierarchical values of the fermion singlet masses correspond to a Majoron field  
peaked on the UV site, with an exponentially decaying profile steeper than the one 
of the ${\rm PS} \to {\rm SM}$ breaking field. 
 As we have shown, the entire flavor structure of the model can be naturally recovered 
 in terms of the single scale ratio in~\eqref{eq:fratios} that controls the relative site separation,  
 and the natural values of bulk masses summarized in~\eqref{eq:bulkM}. 

While the low-energy structure of the 4D model and the 5D continuous one 
are identical, their UV spectra start to differ already around 10 TeV,
given the infinite towers of states appearing in the continuous case. 
In the latter framework, a consistent gauge sector featuring a TeV-scale LQ field able to address the $B$-physics anomalies
clearly points toward a warped geometry, with a warp factor $kL \approx 9$. 
A particularly nice feature of the warped geometry is the emergence of an approximate  
$U(2)^5$ flavor symmetry at the TeV scale, as result of the flatness of the Kaluza-Klein excitations in the UV.
In its minimal implementation, the gauge sector of this 5D construction present some tensions with  
precision measurements of  $Z$-  and $W$-boson couplings to third generation fermions, pointing to non-minimal extensions in the EW sector that we have briefly outlined here
and will be investigated in future work. 

In summary, the warped 5D construction outlined in this paper provides a very promising UV completion
for the flavor non-universal 4321 models addressing the $B$-physics 
anomalies~\cite{Greljo:2018tuh,DiLuzio:2018zxy,Cornella:2019hct,Fuentes-Martin:2019ign,Fuentes-Martin:2020luw,Fuentes-Martin:2020hvc},
while also satisfying the other two foundational aspects of the original PS$^3$ construction, namely the PS-like unification for all the families of quarks and leptons and 
a natural description of the observed fermion spectrum.

\section*{Acknowledgements}
This project has received funding from the European Research Council (ERC) under the European Union's Horizon 2020 research and innovation programme under grant agreement 833280 (FLAY), and by the Swiss National Science Foundation (SNF) under contract 200021-175940. The work of J.F. was supported by the Cluster of Excellence `Precision Physics, Fundamental Interactions, and Structure of Matter' (PRISMA+ EXC 2118/1) funded by the German Research Foundation (DFG) within the German Excellence Strategy (Project ID 39083149).

\appendix

\section{Details on neutrino mass diagonalization}\label{app:NeutrinoDetails}
After spontaneous symmetry breaking, the Lagrangians in \eqref{eq:YukDiag}, \eqref{eq:YukMix}, \eqref{eq:NeutrinoDiag}, and \eqref{eq:NeutrinoMix} yield the following neutrino mass term in the $n_L=(\nu_L \; \nu_R^c \; S_L)^\intercal$ basis
\begin{align} \label{eq:fullnumatrix}
\begin{aligned}
-\mathcal{L_\nu}&\supset\frac{1}{2}~ \overline n_L\,M_\nu\,n_L^c
+ \text{h.c.} \\
&=\frac{1}{2}~ \overline n_L 
\begin{pmatrix}
 0 & M_D \\
M_D^\intercal &  M_R
\end{pmatrix}
n_L^c
+ \text{h.c.}\,,
\end{aligned}
\end{align}
where we defined the block matrices
\begin{align}
M_D	&=\begin{pmatrix}m_D& 0\end{pmatrix}\,,&
M_R &=\begin{pmatrix}0& m_R^\intercal \\ m_R & \mu\end{pmatrix}\,,
\end{align}
with $m_D$, $m_R$ and $\mu$ as in \eqref{eq:massMats}. The mass matrix $M_\nu$ is block-diagonalized by the unitary transformation
\begin{equation} \label{eq:blockdiag}
W^\intercal
\begin{pmatrix}
0 & M_D \\
M_D^\intercal &  M_R
\end{pmatrix} W =    
\begin{pmatrix}
m_\nu & 0 \\
0 &  m_h
\end{pmatrix}
\,,
\end{equation}
where $m_\nu$ and $m_h$ are, respectively, the $3\times3$ light active neutrino and the $6\times 6$ heavy neutrino mass matrices, and the unitary matrix $W$ is parametrized as 
\begin{align}
W = 
\begin{pmatrix}
\sqrt{1-BB^\dagger} & B \\
-B^\dagger & \sqrt{1-B^\dagger B} 
\end{pmatrix}
\,.
\end{align}
Since $\det(M_R)=-\det(m_R^\intercal\, m_R)$, all the eigenvalues of $M_R$ are larger than those of $M_D$, independently of the value of $\mu$, and the block diagonalization in~\eqref{eq:blockdiag} can be performed perturbatively in the $m_D^{\scriptscriptstyle(i)}/m_R^{\scriptscriptstyle(i)}$ expansion~\cite{Schechter:1981cv,Grimus:2000vj,Hettmansperger:2011bt,Dias:2012xp}. At lowest order, we have
\begin{align}
\begin{aligned}
m_\nu &\approx - M_D\, M_R^{-1}  M_D^\intercal \,, \quad m_h\approx M_R\,, \\
B&\approx M_D\, M_R^{-1} \,,
\end{aligned}
\end{align}
with $ M_R^{-1}$ taking the following exact form
\begin{align}
M_R^{-1} = \begin{pmatrix}
-m_R^{-1}\,\mu\,(m_R^\intercal)^{-1} & m_R^{-1} \\[5pt]
(m_R^\intercal)^{-1} & 0
\end{pmatrix}
\,.
\end{align}
Using this expression, we arrive to the final form of the light neutrino mass matrix
 \begin{align} \label{eq:Mlight}
 m_\nu \approx m_D\,  m_R^{-1}\, \mu\, (m_R^\intercal)^{-1}\, m_D^\intercal\,.
 \end{align}
Finally, due to the mixing between active neutrino and sterile states, controlled by $B$, the PMNS matrix $N$ is not unitary anymore. Namely, we have
 \begin{align}\label{eq:UPMNS}
 N \approx \left(1- B B^\dagger/2 \right) U_\ell^T\,U_\nu\,,
 \end{align}
 where $U_\nu$ is the matrix that diagonalize $m_\nu$ and $U_\ell$ the left-handed rotation that brings the charged leptons to their mass eigenbasis. The PMNS non-unitarity is parametrized by the $3 \times 3$ hermitian matrix $\eta  \equiv |\mathbb{1}- N^\dagger N| \approx |B B^\dagger|$, whose expression in terms of $m_D$ and $m_R$ is
 \begin{align} \label{eq:Mlight}
 \eta \approx m_D\,  m_R^{-1} (m_D\, m_R^{-1})^\dagger\,,
 \end{align}
 up to corrections of $\mathcal{O}\big[(m_D^{\scriptscriptstyle(i)}/m_R^{\scriptscriptstyle(i)})^4\big]$.
  
\bibliographystyle{JHEP}
\bibliography{references}

\end{document}